\documentclass[11pt]{JHEP3}

\usepackage{epsfig}

\makeatletter\@addtoreset{equation}{section}\makeatother

\makeatletter

\@addtoreset{equation}{section}
\makeatother

\newcommand{\be}{\begin{equation}}

\newcommand{\ee}{\end{equation}}

\newcommand{\bea}{\begin{eqnarray}}

\newcommand{\eea}{\end{eqnarray}}

\newcommand{\xx}{\nonumber\\}

\newcommand{\ct}{\cite}

\newcommand{\la}{\label}

\newcommand{\eq}[1]{(\ref{#1})}

\def\CN{{\cal N}}

\def\IR{{\hbox{{\rm I}\kern-.2em\hbox{\rm R}}}}
\def\IB{{\hbox{{\rm I}\kern-.2em\hbox{\rm B}}}}
\def\IN{{\hbox{{\rm I}\kern-.2em\hbox{\rm N}}}}
\def\IC{\,\,{\hbox{{\rm I}\kern-.59em\hbox{\bf C}}}}
\def\IZ{{\hbox{{\rm Z}\kern-.4em\hbox{\rm Z}}}}
\def\IP{{\hbox{{\rm I}\kern-.2em\hbox{\rm P}}}}
\def\IH{{\hbox{{\rm I}\kern-.4em\hbox{\rm H}}}}
\def\ID{{\hbox{{\rm I}\kern-.2em\hbox{\rm D}}}}

\def\a{\alpha}

\def\ga{\gamma}

\def\th{\theta}
\def\rh{\rho}
\def\s{\sigma}
\def\t{\tau}

\def\O{\Omega}
\def\G{\Gamma}
\def\S{\Sigma}

\def\half{\frac{1}{2}}

\def\goto{\rightarrow}

\def\pa{\partial}
\def\e0{\epsilon_0}

\def\pr{r^{\prime}}
\def\ps{s^{\prime}}

\def\ad{\dot{a}}

\def\rd{\dot{r}}

\def\rdp{\dot{r}^\prime}
\def\sdp{\dot{s}^\prime}
\def\tdp{\dot{t}^\prime}
\def\rh{\hat{r}}
\def\sh{\hat{s}}
\def\th{\hat{t}}
\def\rhp{\hat{r}^\prime}
\def\shp{\hat{s}^\prime}
\def\thp{\hat{t}^\prime}

\def\so{SO(4)}
\def\sop{SO(4)^\prime}

\preprint{HU-EP-05/25 \\
SOGANG-HEP 314/05 \\
{\tt hep-th/0506091}}

\author{Bum-Hoon Lee, Jong-won Lee and Chanyong Park \\
Department of Physics, Sogang University, Seoul 121-742, Korea \\
E-mail: \email{bhl@ccs.sogang.ac.kr}, \email{yijwon@sogang.ac.kr},
\email{cyong21@ihanyang.ac.kr}}

\author{Hyun Seok Yang \\
Institut f\"ur Physik, Humboldt Universit\"at zu Berlin \\
Newtonstra\ss e 15, D-12489 Berlin, Germany \\
E-mail: \email{hsyang@physik.hu-berlin.de}}

\title{More on supersymmetric D-branes in type IIB plane wave background}

\abstract{We extend the study on D-branes in the type IIB
plane wave background to less supersymmetric configurations. We
show that many new supersymmetric D-branes can be found by turning
on electric as well as magnetic background fluxes, or constantly
boosting D-branes.}


\begin{document}

\renewcommand{\thefootnote}{\arabic{footnote}}
\setcounter{footnote}{0}

\section{Introduction}

It has been known that the type IIB string theory admits three
maximally supersymmetric backgrounds: flat Minkowski space, $AdS_5 \times S^5$
and gravitational plane wave. The last one was recently known with the
discovery \ct{blau} that the type IIB supergravity solution of a gravitational
plane wave with a constant, null five-form field strength
constitutes a maximally supersymmetric background:
\bea \la{pp-metric}
&& ds^2 = -2 dx^+ dx^- - \mu^2 x_I^2 (dx^+)^2 + {dx_I}^2, \\
&& F_{+1234}=F_{+5678}=2 \mu. \nonumber
\eea
The plane wave geometry \eq{pp-metric} is obtained
by taking the Penrose limit of the $AdS_5 \times S^5$ geometry.

The $AdS_5 \times S^5$ geometry got a prominent position due to the AdS/CFT
duality \ct{ads/cft1,ads/cft2,ads/cft3} asserting that the type IIB superstring moving
in the $AdS_5 \times S^5$ background is dual to the four dimensional
$\CN = 4$ super Yang-Mills theory. Though a remarkable success of this
conjecture, a general proof is still out of reach since the string theory in
this background is given by a highly nonlinear two dimensional field theory
\ct{ads5s5} and the duality relates the weak coupling regime of one theory
to the strong coupling regime of the other theory.
Since the plane wave geometry \eq{pp-metric} is obtained through a limit of
the $AdS_5 \times S^5$ geometry, this limit is particularly interesting by
virtue of the AdS/CFT duality. It was realized in \ct{bmn}
that the type IIB string theory in the plane wave background
\eq{pp-metric} has a very simple description in terms of the dual
supersymmetric Yang-Mills theory in a particular double scaling
limit. Remarkably the duality turns out be perturbatively
accessible from both sides of the correspondence, which so truly
goes to a regime of interacting string
theory. For reviews on this subject, see, for example, \ct{bmnreview}.

This kind of concrete realization of the duality is mainly due to the fact that
the string theory in the Ramond-Ramond background \eq{pp-metric}
is exactly solvable \ct{metsaev1,metsaev2}. The plane wave superstring
reduces to a free, massive two dimensional model once one goes to
the light-cone gauge. It is therefore as straightforwardly
quantized as the superstring in a flat spacetime background. It
may thus be possible to get the complete spectrum of the plane
wave superstring including D-branes too.

Since D-branes play a very crucial role in the understanding of string
dualities, AdS/CFT duality, microscopic description of some black-hole
entropy, phenomenological model building, etc. \ct{polchinski,johnson-book},
it is important to have a complete classification of D-branes.
In a flat spacetime, the D-branes appear as the half BPS solitons
of the type II string theories, preserving 16 supersymmetries
and their transverse positions can be arbitrary so that
they are usually the moduli of the BPS solitons.
D-branes can also be described by the boundary states of closed strings
\ct{li,gg}. The symmetries that the boundary state preserves are thus
generically the combination of the closed string symmetries that
leave the boundary state invariant. In this scheme, a D-brane acts
as a source of closed strings and such properties are guaranteed
by the conformal symmetry of the worldsheet.

Though the properties of D-branes have been extensively studied over a decade,
it is still a challenging problem to completely classify the D-branes
in a general string background. Since the string propagation on the curved
background \eq{pp-metric} can be solved exactly by choosing light-cone gauge
in the Green-Schwarz action \ct{metsaev1,metsaev2}, we may have a systematic
classification of D-branes in the type IIB plane wave background.
Two of us with Cha showed in \ct{jhep} that it is actually possible at least
for longitudinal branes, i.e., extended along the light-cone
directions. In this paper we will extend the previous work \ct{jhep}
to less supersymmetric configurations by introducing magnetic as
well as electric fluxes, or constantly boosting D-branes. We will
find a very rich spectrum of supersymmetric D-branes in the type
IIB plane wave background. D-branes in the Ramond-Ramond background \eq{pp-metric}
have been studied in a number of papers \ct{jhep}-\ct{cpark} from different points of view.
Branes in other plane wave backgrounds have also been studied \ct{bak}-\ct{bkk}.

This paper is organized as follows. In section 2 we review the
D-brane classification in \ct{jhep}. In section 3 we start with
eq.\eq{ds-current} derived in \ct{jhep}, which is the most general worldsheet
supercurrent for the open string dynamical supersymmetry applicable to D-branes
with an electric flux $F_{+I}$ and an angular momentum $L_{IJ}$, i.e., boosted
D-branes. We find all possible configurations preserving some fraction of
dynamical supersymmetries for the $D_{\pm}$-branes \ct{billo}-\ct{kim}
and the oblique D-branes (OD-branes) \ct{hikida,ggsns,zamaklar,jhep}.
We will not only reproduce the known solutions but also find many
new supersymmetric D-branes. In section 4 we consider D-branes in
a magnetic flux background $F_{IJ}$, which was recently analyzed by Mattik \ct{matt}
for maximally supersymmetric D-branes. We derive the most general worldsheet
supercurrent \eq{magcurrent} with the magnetic flux $F_{IJ}$. We find several
supersymmetric D-branes, of course, with reproducing the maximally
supersymmetric cases found by Mattik \ct{matt}. In section 5
we study D-branes in the most general background, say, with $F_{+I},
\; L_{IJ}$ as well as $F_{IJ}$. We find there still exist some supersymmetric
D-branes even in this case. In section 6 we briefly review our results
obtained and discuss some related issues. Finally, in appendix A we list
useful matrix relations which are used to find supersymmetric backgrounds
for the oblique D-branes.

\section{D-branes in a plane wave background}

The Green-Schwarz light-cone action in the plane wave background \eq{pp-metric}
describes eight free massive bosons and fermions \ct{metsaev1,metsaev2}. In the
light-cone gauge, $X^+=\tau$, the action is given by
\be \la{gs-action}
S = \frac{1}{2\pi \a^\prime p^+} \int d\tau \int_{0}^{2\pi \a^\prime |p^+|}
d\sigma \Bigl[ \half \pa_+X_I \pa_-X_I - \half \mu^2 X_I^2
- i \bar{S}(\rho^A \pa_A - \mu \Pi)S \Bigr]
\ee
where $\pa_\pm = \pa_\tau \pm \pa_\sigma$.
The equations of motion following from the action \eq{gs-action}
take the form
\bea \la{eom-boson}
&& \pa_+\pa_-X^I + \mu^2 X^I =0, \\
\la{eom-fermion}
&& \pa_+ S^1 - \mu \Pi S^2 =0, \qquad \pa_-S^2 + \mu \Pi S^1 =0.
\eea
The closed string action \eq{gs-action} has the global symmetry
$\so \times \sop \times {\bf Z}_2$ which is the isometry
in the plane wave background \eq{pp-metric}.
The ${\bf Z}_2 $ symmetry here interchanges simultaneously
the two $SO(4)$ directions \ct{chu1}\footnote{\la{z2-symmetry}
This ${\bf Z}_2 $ symmetry explains why the oblique D-brane,
which is at $45^o$ angle in the two $SO(4)$ directions, can
exist in the plane wave background \eq{pp-metric}
and the spectrum of D-branes is symmetric under the ${\bf Z}_2 $ involution \eq{z2}.}
\be \la{z2}
{\bf Z}_2  : (x^1,x^2,x^3,x^4) \leftrightarrow (x^5,x^6,x^7,x^8).
\ee

In this paper we want to study supersymmetric D-branes
in the plane wave background \eq{pp-metric}.
One of doing this is, according to Polchinski \ct{polchinski,johnson-book},
to consider an open string attached on a $Dp$-brane.
The open string theory is then defined by the action \eq{gs-action}
with appropriate boundary conditions imposed on each end of the
open string. So our interest is to find what boundary condition
has to be imposed to preserve (dynamical) supersymmetries in the
open string theory on the $Dp$-brane.
Following the recipe in \ct{jhep}, we will present an efficient worldsheet
formalism for the supersymmetric boundary conditions on the most general ground.

Since the boundary condition of fermionic coordinates is
insensitive to the details of bosonic boundary conditions, we
assume the following boundary condition at each end of the open
string \ct{lambert} without loss of generality:
\be \la{bc-fermion}
(S^1-\Omega S^2)|_{\pa\Sigma}=0,
\ee
where $\Omega$ is a fermionic gluing matrix whose explicit form will be specified.

For D-branes with the flux $F_{+I}$ and the angular momentum
$L_{IJ}$ only, the gluing
matrix $\O$ is exactly the same as the trivial backgrounds and
is simply given by the product of $\ga$-matrices along
the Neumann directions:\footnote{\la{notation}In this paper we will
use the notation and the convention in \ct{jhep}.
Neumann (N) coordinates $X^r$ are decomposed into oblique
directions $X^{\rh}$ and usual parallel directions $X^{\rd}: \; r=(\rh,\rd)$.
Similarly, Dirichlet (D) coordinates $X^{r^\prime}$ are also decomposed into
oblique directions $X^{\rhp}$ and usual parallel directions $X^{\rdp}: \;
r^\prime=(\rhp,\rdp)$.}
\be \la{Omega}
\O = \prod_{r \in N} \ga^r.
\ee
So, in this case,
\bea \la{osquare}
&& \O^2 = \pm 1, \\
\la{anticomm-gaom}
&& \ga^r \O = - \O \ga^r, \;\;\; \forall r \in N, \\
\la{comm-gaom}
&& \ga^{r^\prime} \O = \O \ga^{r^\prime}, \;\;\;\; \forall r^\prime \in D.
\eea

For D-branes with the flux $F_{rs}$, however, $\O$ has the
following form \ct{gg,matt}:
\be \la{magOmega}
\O = \widetilde{\O}\exp^{\frac{1}{4} \Theta_{rs} \ga^{rs}},
\ee
where $\widetilde{\O}$ is the gluing matrix of the type \eq{Omega}
for the Neumann directions without flux
and the parameters $\Theta_{rs}$ depend on the flux $F_{rs}$.
In this case, the nice properties,
eqs.\eq{osquare} and \eq{anticomm-gaom}, no longer hold
due to the additional exponential factor. However the property
\eq{comm-gaom} is still true since the flux $F_{rs}$ extends only
along the Neumann directions. If $\Theta_{rs} \neq 0$, e.g., with rank 2,
the gluing matrix $\O$ continuously interpolates
among codimension 2 D-branes.
When $\Theta_{rs} \goto 0\; \mbox{or} \; \pi$, we have to recover the case
\eq{Omega} \ct{matt}.

As was shown in \ct{jhep}, the possible type of the D-branes with the gluing matrix $\O$ in
eq.\eq{Omega} can be characterized by the matrix $\Gamma$ defined by
\be \la{Gamma}
\Gamma \equiv \Pi\Omega\Pi\Omega.
\ee
It is easy to show that the matrix
$\Gamma$ satisfies the following relations:
\bea \la{Gamma=Gamma}
&& \Pi \Omega \Pi \Omega = \Gamma = \Pi \Omega^T \Pi \Omega^T, \\
\la{Gamma-prop}
&& \Gamma \Gamma^T =1, \qquad \Pi \Gamma \Pi \Gamma = 1.
\eea

The matrix $\Gamma$ is either a symmetric or an antisymmetric matrix.
In the case the matrix $\Gamma$ is symmetric, i.e. $\Gamma^T = \Gamma$,
it follows from \eq{Gamma=Gamma} and \eq{Gamma-prop} that
\be \la{symm-Gamma}
\Gamma^2=1, \qquad [\Pi, \Gamma]=0=[\Omega, \Gamma].
\ee
On the other hand, in the case the matrix $\Gamma$ is antisymmetric, i.e.
$\Gamma^T = - \Gamma$,
\be \la{antisymm-Gamma}
\Gamma^2= - 1, \qquad \{\Pi, \Gamma\}=0=\{\Omega, \Gamma\}.
\ee
It was shown in \ct{jhep} that the D-branes satisfying $\Gamma^2 = -1$ preserve
no supersymmetry. This fact is not affected by introducing
nontrivial backgrounds since the gluing matrix $\O$ is still the same as
before and the matrix $\Gamma$ then has imaginary eigenvalues.
Thus we will consider only the D-branes satisfying $\Gamma^2 = 1$.
Table \ref{tableone} shows the possible D-branes with particular polarizations.
Other D-branes with different polarizations can be obtained
by the $\so \times \sop$ rotations.

\begin{table}[tbp]
\begin{center}
\begin{tabular}{|c|c|c|c|c|} \hline
D-brane type & $\Gamma$         & $\Omega$ \\
\hline
$D_{\pm}$    & $\pm1$           & $\Omega_{D_\pm}$ \\
\hline
$OD3$        & $\pm \gamma^{1256}$ & $\frac{1}{2}(\gamma^{1}\!-\!\gamma^{6})
(\gamma^{2}\! \pm \!\gamma^{5})$ \\
\hline
             &                  & $\frac{1}{2}(\gamma^{1}\!-\!\gamma^{6})
             (\gamma^{2}\! \mp \!\gamma^{5})\gamma^{34}$ \\
$OD5$        & $\pm \gamma^{1256}$ & $\frac{1}{2}(\gamma^{1}\!-\!\gamma^{6})
(\gamma^{2}\! \mp \!\gamma^{5}) \gamma^{78}$ \\
             &                  & $\frac{1}{2}(\gamma^{1}\!-\!\gamma^{6})
             (\gamma^{2}\! \pm \!\gamma^{5})\gamma^{37}$ \\
\hline
$OD7$        & $\pm \gamma^{1256}$ & $\frac{1}{2}(\gamma^{1}\!-\!\gamma^{6})
             (\gamma^{2}\! \pm \!\gamma^{5})\gamma^{3478}$ \\
\hline
$OD_{\pm}5$  & $\pm\gamma$    & $\frac{1}{4}(\gamma^{1}\!-\!\gamma^{6})
(\gamma^{2}\! \pm \!\gamma^{5}) (\gamma^{3}\!-\!\gamma^{8})(\gamma^{4}\!+\!\gamma^{7})$ \\
\hline
\end{tabular}
\end{center}
\caption{D-branes with $\Gamma^2 = 1$}
\label{tableone}
\end{table}

$D_\pm$-branes \ct{billo}-\ct{kim} are a specific class
satisfying $\Gamma = \pm 1$, which are denoted as $(+,-,m,n)$ with $m,n=0,1,\cdots,4$
following the convention in \ct{sken-tayl}.
$D_-$-branes are of the type $|m-n|= 2$ while $D_+$-branes are of
the type $|m-n|= 0,4$. The oblique D-branes with $\Gamma^2 = 1$ can be summarized as follows:
\begin{eqnarray}
\label{2-Gamma}
&& \mbox{$ODp$-brane}: \Gamma = \pm \gamma^{i_1 i_2 i^\prime_3
i^\prime_4}, \quad (p=3,5,7), \\
\label{4-Gamma}
&& \mbox{$OD5$-brane}: \Gamma = \pm \gamma,
\end{eqnarray}
where $\ga = \ga^{12\cdots 8}$ is the $SO(8)$ chirality matrix.
Eq.\eq{symm-Gamma} requires that
$\Gamma$ should contain an even number of gamma matrices
in both $\{ \gamma^i, \; i=1, \cdots, 4 \}$ and $\{ \gamma^{i^\prime},
\; i^\prime=5, \cdots, 8 \}$.

\section{Supersymmetric D-branes with $F_{+I}$ and $L_{IJ}$}

In a light-cone gauge, the 32 components of the supersymmetries
for a closed string decompose into kinematical supercharges,
$Q^{+A}_a$, and dynamical supercharges, $Q^{-A}_{\ad}$.
For a closed superstring in the plane wave background with
the action \eq{gs-action}, the conserved super-N\"other
charges were identified by Metsaev \ct{metsaev1}:
\bea \la{charge-p}
&& Q^{+1} = \frac{\sqrt{2p^+}}{2\pi \a^\prime p^+}
\int_{0}^{2\pi \a^\prime |p^+|} d\sigma (\cos\mu\tau S^1 -\sin \mu\tau \Pi S^2), \\
\la{charge-q+2}
&& Q^{+2} = \frac{\sqrt{2p^+}}{2\pi \a^\prime p^+}
\int_{0}^{2\pi \a^\prime |p^+|} d\sigma (\cos\mu\tau S^2 + \sin \mu\tau \Pi S^1), \\
\la{charge-q-1}
&& \sqrt{2p^+}Q^{-1}= \frac{1}{2\pi \a^\prime p^+}
\int_{0}^{2\pi \a^\prime |p^+|} d\sigma \Bigl( \pa_- X^I\gamma^I S^1
-\mu X^I\gamma^I \Pi S^2 \Bigr), \\
\la{charge-q-2}
&& \sqrt{2p^+}Q^{-2}= \frac{1}{2\pi \a^\prime p^+}
\int_{0}^{2\pi \a^\prime |p^+|} d\sigma \Bigl( \pa_+ X^I\gamma^I
S^2 + \mu X^I\gamma^I \Pi S^1 \Bigr).
\eea

The kinematical supersymmetry is, in general, related to a shift
of spinor fields and thus generated by spinor zero modes. So the
kinematical supersymmetry is insensitive to the details of
backgrounds, i.e., fluxes and boosting,\footnote{But the explicit form
of the spinor zero modes themselves is sensitive to the type of
D-brane and the background gauge condensates \ct{jhep,matt}.}
and it has to be fixed by the boundary condition
\eq{bc-fermion}. Since we are interested in the open string
supersymmetry surviving nontrivial backgrounds, we will focus only
on the dynamical supersymmetry. The dynamical supercharge
preserved by an open string on a D-brane is given by a combination
of closed string supercharges $Q^{-A}$ compatible with the open
string boundary conditions.
Due to the boundary condition \eq{bc-fermion},
it turns out that the conserved dynamical supercharge is given by
(a subset of)
\be \la{dyn-susy}
q^- = Q^{-1}- \Omega Q^{-2}.
\ee

In this section we will first show how D-branes can preserve
dynamical supersymmetries by turning on the flux $F_{+I}$ or the
angular momentum $L_{IJ}$. It is easy to derive the
conservation law \ct{jhep} for the dynamical supersymmetry in eq.\eq{dyn-susy}
using the equations of motion, eqs. \eq{eom-boson} and
\eq{eom-fermion}:
\be \la{con-law-dyn}
\frac{\pa q^-_\tau}{\pa \tau} + \frac{\pa q^-_\sigma}
{\pa \sigma} = 0,
\ee
where
\bea \la{ds-current}
q^-_\sigma &=& \sqrt{\frac{1}{2p^+}}
\Bigl( (\pa_\tau X^r \gamma^r - \pa_\sigma X^{r^\prime} \gamma^{r^\prime})
(S^1 - \Omega S^2) \xx
&& + (\pa_\tau X^{r^\prime} \gamma^{r^\prime} - \pa_\sigma X^r \gamma^r)
(S^1 + \Omega S^2) \xx
&& + \mu X^r \gamma^r \Omega \Pi (S^1 + \Gamma \Omega S^2)
- \mu X^{r^\prime} \gamma^{r^\prime} \Omega \Pi (S^1 - \Gamma
\Omega  S^2) \Bigr).
\eea
In the course of derivation, we used the relations, \eq{anticomm-gaom} and
\eq{comm-gaom}. However, we didn't assume anything about bosonic as well as
fermionic boundary conditions.

In order for the supercharge $q^-$ to be conserved,
the current $q^-_\sigma$ in eq.\eq{ds-current} has to vanish
at the boundary of worldsheet, $\partial \Sigma$.
Now we assume the fermionic boundary condition \eq{bc-fermion},
but it does not loose any generality since (the form of) the boundary
condition \eq{bc-fermion} does not depend on the details of backgrounds.
Then we will find what boundary conditions for bosonic coordinates $X^I$
have to be imposed to get the vanishing current at the boundary,
i.e., $q^-_\sigma|_{\partial \Sigma} = 0$.
Since the details of the bosonic boundary condition, however,
depend on the type of D-brane, we will discuss $D_{\pm}$-branes
and $OD$-branes, separately.

\subsection{$D_{-}$-branes}

First we consider the dynamical supersymmetry of $D_-$-branes,
where $\G = -1$.  In this case, the current $q^-_\sigma$ in eq. \eq{ds-current} at the
boundary reduces to
\be \la{d-current}
q_{\s}^- |_{\pa \S} = \sqrt{\frac{2}{p^+}} \Bigl(\pa_{\t} X^{\pr} \ga^{\pr}
- \pa_{\s} X^r \ga^r
- \mu X^{\pr} \ga^{\pr} \O \Pi \Bigr) S^1 |_{\pa \S}.
\ee
We want to find what conditions are needed for bosonic coordinates $X^I$
in order for the current \eq{d-current} to vanish at the boundary.
Of course, the trivial case is
\be \la{d-trivial}
\pa_{\t} X^{\pr} =  \pa_{\s} X^r =  X^{\pr} = 0,
\;\;\; \forall \pr \in D, \; \forall r \in N,
\ee
and this configuration preserves maximal supersymmetry.
But, as we will discuss, there are many other configurations with the vanishing current
which so preserve some amount of supersymmetries.

It is useful to notice that the matrix $\O \Pi$ for the $D_-$-branes takes the following form:
\be \la{matrix-op}
\O \Pi = \pm \ga^{I_1 I_2} \;\; \mbox{or} \;\; \pm \ga \ga^{I_1 I_2},
\ee
where
\bea \la{matrix-d-3}
&& \mbox{$D_-3$-brane}: \; (I_1, I_2) \in  D, \\
\la{matrix-d-35}
&& \mbox{$D_-5$-brane}: \; I_1 \in  N, \;\;  I_2 \in  D, \\
\la{matrix-d-7}
&& \mbox{$D_-7$-brane}: \; (I_1, I_2) \in  N.
\eea

\subsubsection{$\pa_{\s} X^{r} \neq 0$ case}

If we consider the case $X^{\pr}|_{\pa \S} \equiv x_0^{\pr} \neq 0$ for some $\pr \in D$,
our problem is reduced to that finding a matrix satisfying
\be \la{d-flux}
\ga^{r \pr} \O \Pi S^1 = \pm S^1.
\ee
A necessary condition is that $(\ga^{r \pr} \O \Pi)^2 = 1$.
Therefore the matrix $\ga^{r \pr} \O \Pi$ has to take the form
\be \la{matrix1}
\ga^{r \pr} \O \Pi = \pm \ga^{I_1 \cdots I_n},
\;\;\; n=0,4,8.
\ee
If a matrix exists satisfying eq.\eq{d-flux}, the current at the boundary can
vanish with the following modified Neumann boundary condition:
\begin{equation}\label{d-flux-bc}
    \pa_\sigma X^r - \mu x_0^{\pr} = 0.
\end{equation}
This kind of boundary condition can be easily achieved by introducing
a boundary coupling with the worldvolume gauge field $A_+ = -F_{+r}X^r$:
\begin{equation}\label{bd-action}
S_B = - \frac{1}{2\pi \a^\prime p^+} \int_{\pa \Sigma} d\tau A_\mu (X)
\frac{\pa X^\mu}{\pa \tau} = \frac{F_{+ r}}{2 \pi \a^\prime p^+}
\int_{\pa \Sigma}  d\tau X^r.
\end{equation}
When the spinor $S^A$ satisfies \eq{d-flux} together with the Neumann boundary
condition \eq{d-flux-bc}, the dynamical supersymmetries
given by $\half( 1 \pm \ga^{r \pr} \O \Pi) q^-$ are preserved.

Going with eq.\eq{matrix-op} into eq.\eq{matrix1}, it is easy to see that
the $n=0$ and $8$ cases are possible only for D5-branes:
$(+,-,3,1)$ and $(+,-,1,3)$ which preserve maximal supersymmetry
as was shown in \ct{sken-tayl,tt,jhep}.
For example, let us take a $(+,-,3,1)$-brane extended along
$(+,-,1,2,3,5)$ directions, say, $N= (1,2,3,5)$ and $ D=
(4,6,7,8)$ and thus $\Omega\Pi = - \gamma^{45}$.
In this case we need the flux $F_{+5} = \mu x_0^4$ only.

Using eq.\eq{matrix-op}, it is obvious that the $n=4$ case is possible for all $D_-$-branes.
Four dynamical supersymmetries are preserved in this case. We will
not give any detail since it should be really simple. Instead let
us give you an example: Consider $(+,-,2,0)$-brane where $\O = \ga^{12}$ and $\O \Pi = - \ga^{34}$.
If $X^5|_{\pa \Sigma} = x_0^5 \neq 0$ and the spinor satisfies
$(1 \pm \ga^{1345}) S^1 =0$ at the boundary,
the half of dynamical supercharges are preserved with the boundary
condition $\pa_\s X^1 \mp \mu x_0^5 =0$.

We can get less supersymmetric configurations by considering more
general backgrounds. For example, let us consider two
matrices $M_1$ and $M_2$ satisfying eq.\eq{d-flux}.
In order for the spinor $S^1$ to simultaneously satisfy the
condition \eq{d-flux} for this background, the product of
$M_1$ and $M_2$, $M_3= M_1 M_2$, should again be of the form
\eq{matrix1}. In a pedantic notation,
\be \la{siminter}
M_1 S^1 = \pm S^1 \; \mbox{and} \; M_2 S^1 = \pm S^1
\; \Rightarrow \; M_3= M_1 M_2 = \pm \ga^{I_1 \cdots I_n},
\;\;\; n=0,4,8.
\ee
If $M_3$ is of the form with $n=0,8$, the supersymmetry is not
further broken. But, the dynamical supersymmetry is further broken
by half in the case of $n=4$.

What is the least supersymmetric configuration which can be
realized by turning on constant fluxes $F_{+I}$ ?
Since $M_3= M_1 M_2$ should be of the form in eq.\eq{siminter},
we can see from eqs.\eq{matrix-d-3}-\eq{matrix-d-7}
that $D3$- and $D7$-branes can have only two independent
projections - 2 dynamical supersymmetries. This can be easily
understood by noting that the
$D3\;(D7)$-brane has only two Neumann (Dirichlet) directions.
For the D5-brane discussed above, for example,
we can have $M_1 = \ga^{1845}$, $M_2 = \ga^{2745}$ and $M_3 =
\ga^{3645}$, but $M_1M_2M_3 = - \ga$, so $M_1, \, M_2$ and $M_3$
cannot be simultaneously independent in the space of positive
chirality spinors. Therefore the D5-brane also preserves at least 2
dynamical supersymmetries.

\subsubsection{$\pa_{\t} X^{\pr} \neq 0$ case}

If we consider the case $X^{\ps}|_{\pa \S} \equiv v^{\ps} \neq 0$ for some
$\ps \in D$, we need a modified Dirichlet boundary condition:
\begin{equation}\label{d-ro-bc}
    \pa_\tau X^{\pr} - \mu v^{\ps} = 0.
\end{equation}
This kind of boundary condition can be achieved by boosting a D-brane with
constant velocity $v^{\ps}$ in a transverse direction.
This means we are considering the following transformation
\be \la{boosting}
  X^{\pr} \rightarrow  X^{\pr} - \mu v^{\ps} \tau
\ee
where the light-cone gauge $X^+ = \tau$ is used.
With the boundary condition \eq{d-ro-bc}, the supersymmetric condition
is reduced to that finding a matrix satisfying
\be \la{d-rotation}
\ga^{\pr \ps} \O \Pi S^1 = \pm S^1.
\ee
Therefore the matrix $\ga^{\pr \ps} \O \Pi$ has to take the form
\be \la{matrix2}
\ga^{\pr \ps} \O \Pi = \pm \ga^{I_1 \cdots I_n},
\;\;\; n=0,4,8.
\ee
Note that $\ga^{\pr \ps}$ is a $SO(2)$ spinor rotation in the transverse
rotational symmetry $SO(4-m) \times SO(4-n)$ for a $(+,-,m,n)$-brane.

Going with eq.\eq{matrix-op} into eq.\eq{matrix2}, it is easy to see that
the $n=0$ and $8$ cases are possible only for D3-branes:
$(+,-,2,0)$ and $(+,-,0,2)$ which preserve maximal supersymmetry
as was shown in \ct{tt}. This is the case that $\ga^{\pr \ps} \in SO(2)$
in the transverse rotation symmetry $SO(4) \times SO(2)$.
However, the $n=4$ case is possible for all
$D_-$-branes in which case four dynamical supersymmetries are preserved.
These branes are rotating in the $X^{\pr}$-$X^{\ps}$ plane and correspond to
the giant gravitons.

We can get less supersymmetric configurations by considering more
boostings. What is the least supersymmetric configuration which can be
realized by boosting a D-brane ? If we consider two boosts simultaneously,
the product of $M_1$ and $M_2$, $M_3= M_1 M_2$, should be of the form
in eq.\eq{matrix2} where the matrices $M_1$ and $M_2$ satisfy eq.\eq{d-rotation}.
Then we can see from eqs.\eq{matrix-d-3}-\eq{matrix-d-7}
that $D5$- and $D7$-branes can have only one supersymmetric
rotation - 4 dynamical supersymmetries. This can be easily
understood by noting that the $SO(3)\; (SO(2))$ rotation
for the $D5 \;(D7)$-brane is rank 1.
For the D3-brane, however, we can have two simultaneous rotations in the
transverse $SO(4)$ directions - 2 dynamical supersymmetries since $SO(4)$ is
rank 2. The simultaneous $SO(2)$ rotation of the D3-brane does not
further break supersymmetry as the reason discussed above.

\subsubsection{general case}

Now we consider general cases with
$\pa_{\t} X^{\pr} \neq 0$ and $\pa_{\s} X^{r} \neq 0$.
In this case we have two kinds of matrix from the conditions
\eq{d-flux} and \eq{d-rotation}. One is of the form $M^F = \ga^{r \pr} \O
\Pi$ and the other is $M^L = \ga^{\ps t^\prime} \O \Pi$.
To preserve the dynamical supersymmetry,
the following condition is further required:
\be \la{d-general}
M^F M^L = \pm \ga^{r \pr\ps t^\prime}.
\ee
Thus we need at least three Dirichlet directions.
Note that we can simply add the maximally supersymmetric configuration in the previous
cases not affecting the resulting supersymmetry
only if the condition \eq{d-general} is satisfied.
So we will discuss supersymmetric configurations up to the
maximally supersymmetric background in 3.1.1 and 3.1.2.

The condition \eq{d-general} says that this case can preserve at most 4
dynamical supersymmetries. It also says that the D7-brane cannot preserve any
dynamical supersymmetry in this case.
Noting that $M^L = \ga^{r \pr \ps t^\prime}$ for the D5-brane,
the background with one flux and one rotation can preserve 2
dynamical supersymmtry as the least supersymmetric configuration.
For the D3-brane, first note that
$M^F = \ga^{r \pr \ps t^\prime}, \;M^L = \ga^{\pr \ps t^\prime u^\prime}$
and so we can have only two independent
projections satisfying the condition $\eq{d-general}$.
For example, for the D3-brane discussed in 3.1.1,
$M_1^F = \ga^{1345}, \; M_2^F = \ga^{2346}$ and  $M_1^L = \ga^{3478}$.
Since $M_1^F M_2^F M_1^L = \ga$,
the dynamical supersymmetry is reduced only by $1/4$.

\subsection{$D_{+}$-branes}

Next we consider the dynamical supersymmetry of $D_+$-branes,
where $\G = +1$.  In this case, the current $q^-_\sigma$ in eq. \eq{ds-current} at the
boundary reduces to
\be \la{d+current}
q_{\s}^- |_{\pa \S} = \sqrt{\frac{2}{p^+}} \Bigl(\pa_{\t} X^{\pr} \ga^{\pr}
- \pa_{\s} X^r \ga^r
+ \mu X^r \ga^{r} \O \Pi \Bigr) S^1 |_{\pa \S}.
\ee
A crucial difference from the $D_-$-branes is that the term proportional to
$\mu$ is now involved with Neumann coordinates, which are in general nonvanishing and
$\t$-dependent at the boundary. So we can realize a supersymmetric
configuration neither by turning on a constant flux nor by boosting the D-brane
unlike as $D_-$-branes.

Nevertheless, as was found in \ct{jhep}, the dynamical
supersymmetry can be preserved by introducing
a boundary coupling with the worldvolume gauge field:
\begin{equation}\label{bd+action}
S_B = - \frac{1}{2\pi \a^\prime p^+} \int_{\pa \Sigma} d\tau A_\mu (X)
\frac{\pa X^\mu}{\pa \tau} = - \frac{1}{2 \pi \a^\prime p^+}
\int_{\pa \Sigma}  d\tau A_+(X),
\end{equation}
where the flux $F_{+I}$ is not constant but linearly depends on
the Neumann coordinates. That is, the gauge field $A_+(X)$ is
given by
\be \la{d+a}
A_+(X) = \pm \frac{\mu}{2} \Bigl( \sum_{r_1 \in N_1} X^{r_1} X^{r_1}
-  \sum_{r_2 \in N_2} X^{r_2} X^{r_2} \Bigr),
\ee
where $N_1$ denotes Neumann coordinates in the first $SO(4)$
directions and $N_2$ does those in the second $SO(4)$ directions.
The Neumann boundary condition is then modified as follows
\be \la{mbc-d+}
\Bigl(\partial_\sigma X^{r_1}
\pm \mu X^{r_1} \Bigr)_{\partial\Sigma} = 0 =
\Bigl(\partial_\sigma X^{r_2}
\mp \mu X^{r_2} \Bigr)_{\partial\Sigma}.
\ee

The dynamical supersymmetry of $D_+$-branes can be preserved
basically due to the fact that $(\O \Pi)^2 =1$ so that there are always solutions
satisfying $\O \Pi S^1 = \pm S^1$. In particular, the $(+,-,4,0)$-
and $(+,-,0,4)$-brane preserve the maximal supersymmetry since $\O \Pi =
1$ and $\ga$, respectively, for these branes \ct{sken-tayl,gaberdiel,kim}.
One may ask whether or not less supersymmetric configurations can
be constructed. Looking into the structure of the current in
eq.\eq{d+current}, it seems to be impossible.

\subsection{$OD$-branes}

According to the gluing matrix $\Omega$ in table \ref{tableone},
we will define diagonal coordinates
\begin{equation}\label{dia-coordinate}
    X^{\rh} = \frac{1}{\sqrt{2}}(X^r \pm X^{r^\prime}),
    \qquad X^{\rhp} = \frac{1}{\sqrt{2}}(X^{r^\prime} \mp X^r)
\end{equation}
with the index notation explained in footnote \ref{notation}.
For an OD5-brane described by $\Omega = \frac{1}{2}(\gamma^{1}\!-\!\gamma^{6})
(\gamma^{2}\! - \!\gamma^{5})\gamma^{34}$, for example, we have
\bea \la{od5-n}
&& \mbox{Neumann}: X^{\hat{1}} = \frac{1}{\sqrt{2}}(X^1 - X^{6}), \;
X^{\hat{2}} = \frac{1}{\sqrt{2}}(X^2 - X^{5}), \;
X^{\dot{3}} = X^3, \; X^{\dot{4}} = X^4, \xx
&& \mbox{Dirichlet}: X^{\hat{5}^\prime} = \frac{1}{\sqrt{2}}(X^5 + X^{2}), \;
X^{\hat{6}^\prime} = \frac{1}{\sqrt{2}}(X^6 + X^{1}), \;
X^{\dot{7}^\prime} = X^7, \; X^{\dot{8}^\prime} = X^8. \nonumber
\eea

To discuss the supersymmetry of $OD$-branes,
it is useful to decompose the spinors $S^A(\tau,\sigma)$ into
the eigenspinors of $\Gamma$ by defining
\be \la{dec-s}
S^A_\pm(\tau,\sigma) = P_\pm S^A(\tau,\sigma),
\ee
where
\begin{equation}\label{projection}
    P_\pm = \half(1 \pm \Gamma).
\end{equation}
It follows from eq.\eq{symm-Gamma} that the equations of motion, eq.\eq{eom-fermion},
are completely separated into two independent equations of motion
for the spinors $S_\pm^A(\tau, \sigma)$
\bea \la{+eom-s+s-}
&& \pa_+ S_+^1 - \mu \Pi S_+^2 =0, \qquad  \pa_- S_+^2 + \mu \Pi S_+^1 =0, \\
\la{-eom-s+s-}
&& \pa_+ S_-^1 -\mu \Pi S_-^2 =0, \qquad \pa_- S_-^2 + \mu \Pi S_-^1=0
\eea
and the boundary condition, eq.\eq{bc-fermion}, can be separately
imposed for the spinors $S_\pm^A(\tau, \sigma)$
\bea \la{bc-s+}
&& (S^1_+ - \Omega S^2_+)|_{\pa\Sigma}=0, \\
\la{bc-s-}
&& (S^1_- - \Omega S^2_-)|_{\pa\Sigma}=0.
\eea
It can be shown \ct{jhep} that the spinor $S_+^A(\tau, \sigma)$ then has a
$D_+$-like mode expansion while $S_-^A(\tau, \sigma)$ does a
$D_-$-like mode expansion since
\be \la{eigen-Gamma}
\Gamma S^A_\pm(\tau,\sigma) = \pm S^A_\pm(\tau,\sigma).
\ee

Since the condition for $q^-_{\sigma}|_{\pa \Sigma}$ in eq.\eq{ds-current}
to vanish depends on the eigenvalue of the matrix $\Gamma$ as was reasoned above,
we introduce projected supercharges defined by
\be \la{projdyn-susy}
q^-_\pm \equiv P_\pm (Q^{-1}- \Omega Q^{-2}).
\ee
It is easy to get the value of the current
$q^-_{\pm \sigma}$ at the boundary:
\begin{eqnarray}\label{dynq+sigma}
q^-_{+\sigma} \bigg|_{\partial\Sigma} = \sqrt{\frac{2}{p^+}}
\Bigl( && (\pa_\t X^{\rdp} \ga^{\rdp} - \partial_\sigma X^{\rd} \gamma^{\rd}
+ \mu X^{\rd} \gamma^{\rd } \Omega \Pi) S^1_+ \xx
+ && (\pa_\t X^{\rhp} \ga^{\rhp} - \partial_\sigma X^{\hat{r}}\gamma^{\hat{r}}
- \mu X^{\rhp}\gamma^{\rhp}\Omega \Pi) S^1_- \Bigr)_{\partial\Sigma}
\end{eqnarray}
and
\begin{eqnarray}\label{dynq-sigma}
q^-_{-\sigma} \bigg|_{\partial\Sigma} = \sqrt{\frac{2}{p^+}}
\Bigl( && (\pa_\t X^{\rhp} \ga^{\rhp} - \partial_\sigma X^{\rh} \gamma^{\rh}
+ \mu X^{\rh} \gamma^{\rh} \Omega \Pi) S^1_+  \xx
+ && (\pa_\t X^{\rdp} \ga^{\rdp} - \partial_\sigma X^{\rd}\gamma^{\rd}
- \mu X^{\rdp}\gamma^{\rdp}\Omega \Pi) S^1_- \Bigr)_{\partial\Sigma}.
\end{eqnarray}
Note that the dynamical supersymmetry, $q^-_+$ and $q^-_-$, cannot
simultaneously be preserved since each set of boundary conditions
cannot simultaneously be compatible with each other. So we will
separately consider the supercharges $q_{\pm}^-$.

We see that the $+$ component of the spinor $S^1$ in
eqs.\eq{dynq+sigma}-\eq{dynq-sigma} gives $D_+$-like supercharge,
while the $-$ component gives $D_-$-like supercharge.
As was shown in \ct{jhep}, the $D_+$-like supercharge can be preserved by turning on
a boundary coupling with the gauge field $A_{+}(X)$ like eq.\eq{d+a}.
One can easily understand the results by looking into the matrix
relations in appendix A. Especially, the $OD3$-brane preserves 4
dynamical supersymmetries $q_+^-$ \ct{hikida,ggsns} since $X^{\rd}=0$ by definition
while it can do only 2 dynamical supersymmetries $q_{-}^-$ by
turning on a boundary coupling with the gauge field $A_{+}(X)$ of the type
\eq{d+a}. The $OD5$-branes with $\O = \frac{1}{2}(\gamma^{1}-\gamma^{6})
(\gamma^{2} - \gamma^{5}) \ga^{34}$ and $\O = \frac{1}{2}(\gamma^{1}-\gamma^{6})
(\gamma^{2} - \gamma^{5}) \ga^{78}$ also preserve $q_{\pm}^-$ without any further
projection since they satisfy eq.\eq{0d5+34} and eq.\eq{0d5+78}, respectively.

We go over to the $-$ component of the spinor $S^1$ which gives
$D_-$-like supercharge. We are now interested in the situation
$X^{\pr} \neq 0$ for some $\pr \in D$. One has to remember that we
already introduced one or two projection operators to preserve the
$D_+$-like supercharge, so that the introduction of nontrivial
backgrounds for the $D_-$-like supercharge may further break the
supersymmetry. It could be helpful to have an analogue of eq.\eq{matrix-op} for the
$OD$-branes. In appendix A we list the useful matrix relations for
those in table \ref{tableone}. The matrix relations show a quite similar
property to eq.\eq{matrix-op} so that we can apply the
same strategy as the $D_-$-branes. For this, we will often use the simple fact, for example,
\be \la{ga+-}
(\gamma^I - \gamma^J)(\gamma^I + \gamma^J) = 2 \gamma^{IJ}.
\ee
We will not repeat how to modify the boundary conditions for the
$OD$-branes by turning on a flux or boosting a D-brane since it is
essentially the same as the $D_-$-branes.

Let us first discuss the $OD_{\pm}5$-branes since they are special compared to
other $OD$-branes. As was shown in \ct{jhep}, the $OD_+5$-brane preserves no
dynamical supersymmetry. For the $OD_-5$-brane, however,
the current $q^-_{+ \sigma}$ at the boundary ($q^-_-$ identically vanishes) is
given by
\be \label{dynod-}
q^-_{+\sigma} \bigg|_{\partial\Sigma} = \sqrt{\frac{2}{p^+}}
(\pa_\t X^{\rhp} \ga^{\rhp} - \partial_\sigma X^{\hat{r}}\gamma^{\hat{r}}
- \mu X^{\rhp}\gamma^{\rhp}\Omega \Pi) S^1 |_{\partial\Sigma}.
\ee
So the trivial boundary condition $\pa_\t X^{\rhp}=\partial_\sigma
X^{\hat{r}}= X^{\rhp}=0$ preserves the maximal supersymmetry \ct{hikida,ggsns}.
Now our question is whether or not some dynamical supersymmetry
can be preserved by introducing a constant flux or a boosting.
The answer is no since $\O$ contains too many (4)
oblique Neumann directions and so the vanishing condition in
eq.\eq{dynod-} can also be involved with the product of 2 or 6 gamma matrices.
In the following we will thus discuss the other $OD$-branes only.

\subsubsection{$\pa_{\s} X^{r} \neq 0$ case}

In this case the problem is to find a matrix satisfying
\begin{eqnarray}\label{odmatrix-flux}
q_+^-: \; \gamma^{\rh\rhp}\Omega \Pi S^1_- = \pm S^1_-
\end{eqnarray}
or
\begin{eqnarray}\label{odmatrix+flux}
q_-^-: \; \gamma^{\rd\rdp}\Omega \Pi S^1_- = \pm S^1_-,
\end{eqnarray}
where the $OD3$-brane can preserve $q_-^-$ with trivial Dirichlet
boundary condition $X^{\rdp}=0$ only since $X^{\rd}=0$ by definition
while the $OD7$-brane can preserve $q_-^-$ with trivial Neumann
boundary condition $\partial_\sigma X^{\rd} = 0$ since $X^{\rdp}=0$.
Otherwise we are implicitly assuming $X^{\pr} \neq 0$ for
the related Dirichlet coordinates. Of course, the supersymmetry $q_{\pm}^-$
can be preserved with the trivial boundary condition
when $X^{\pr} = 0$, which is not the case of our interest.
For the gamma matrices in eq.\eq{odmatrix-flux},
it is convenient to distinguish the following two cases
\bea \la{odga-con}
&& \ga^{\rhp} = \pm \Pi \ga^{\rh} \Pi, \\
\la{odga-notcon}
&& \ga^{\rhp} \neq \pm \Pi \ga^{\rh} \Pi,
\eea
since their supersymmetry will be different in general.

From eqs.\eq{0d5-34} and \eq{0d5-78}, we see that the
corresponding $OD5$-branes preserve 4 dynamical supersymmetries
$q^-_+$ for the case \eq{odga-con} while no supersymmetry for the
case \eq{odga-notcon}. For example, the $OD5$-brane with
$\O = \frac{1}{2}(\gamma^{1}-\gamma^{6})
(\gamma^{2} - \gamma^{5}) \ga^{34}$ has the value $\gamma^{\rh\rhp}\Omega \Pi
= -\ga^{1634}\ga^{1234} = \ga^{26}$ for the case
\eq{odga-notcon}. The other $OD$-branes preserve 2 dynamical
supersymmetries $q^-_+$ for the case \eq{odga-con}
since we meet again the same projection operators as those in
eqs.\eq{0d3+}, \eq{0d5+37} and \eq{0d7+}.
However the case \eq{odga-notcon} cannot preserve the supersymmetry $q^-_+$
except the $OD3$-brane
since the projection operators for the $D_-$-like supercharge are
not compatible with those for the $D_+$-like supercharge.
For example, the $OD5$-brane with $\O = \frac{1}{2}(\gamma^{1}-\gamma^{6})
(\gamma^{2} + \gamma^{5}) \ga^{37}$ has a value $\gamma^{\rh\rhp}\Omega \Pi
= \ga^{2537}\ga^{1234} = \ga^{1457}$ for the case \eq{odga-notcon}
whose product with the matrices in eq.\eq{0d5+37} becomes $\ga^{12}$ or $\ga^{56}$.
The $OD3$-brane did not yet use the matrices in eq.\eq{0d3+} for the
$D_+$-like supercharge, so it preserves 2 dynamical
supersymmetries $q^-_+$ even for the case \eq{odga-notcon}.
On the other hand, we see that all the $OD5$-branes preserve 2
dynamical supersymmetries $q^-_-$ since totally two independent
projections are needed.

\subsubsection{$\pa_{\t} X^{\pr} \neq 0$ case}

In this case we have a condition
\begin{eqnarray}\label{odmatrix-rot}
q_+^-: \; \gamma^{\rhp \shp}\Omega \Pi S^1_- = \pm S^1_-
\end{eqnarray}
or
\begin{eqnarray}\label{odmatrix+rot}
q_-^-: \; \gamma^{\rdp\sdp}\Omega \Pi S^1_- = \pm S^1_-,
\end{eqnarray}
where the $OD7$-brane does not belong to the case \eq{odmatrix+rot} since
$X^{\rdp} = 0$ by definition.

Noting that $\gamma^{\rhp \shp} \gamma^{\rh \sh} = \pm
\ga^{1256}$ for the table \ref{tableone} and following the similar reasoning to 3.3.1,
we immediately see that only the $OD3$-brane preserves 2 dynamical
supersymmetries $q^-_+$. Using the relations in appendix A, we
also easily see that the $OD3$-brane preserves 2 dynamical
supersymmetries $q^-_-$ since no further projection is needed and
the $OD5$-branes also do 2 dynamical supersymmetries except the
$OD5$-brane with $\O = \frac{1}{2}(\gamma^{1}-\gamma^{6})(\gamma^{2} - \gamma^{5})
\ga^{37}$ which preserves no supersymmetry $q^-_-$
unless $X^{\rdp}=0$.

\subsubsection{general case}

Finally we consider the general case with $\pa_{\t} X^{\pr} \neq 0$
and $\pa_{\s} X^{r} \neq 0$.
Since the conditions \eq{odmatrix-flux} and \eq{odmatrix-rot} or
\eq{odmatrix+flux} and \eq{odmatrix+rot} have to be simultaneously
satisfied, we have an additional condition as in 3.1.3 coming from
the product
\bea \la{odgen-mat}
&& q_+^-: \; \gamma^{\rh\rhp}\Omega \Pi \gamma^{\shp \thp}\Omega
\Pi = \ga^{\rh \sh \th \rhp} \G, \\
\la{odgen+mat}
&& q_-^-: \; \gamma^{\rd\rdp}\Omega \Pi \gamma^{\sdp\tdp}\Omega
\Pi = \pm \ga^{\rd \rdp \sdp \tdp} \G,
\eea
where eq.\eq{odga-con} has been used in eq.\eq{odgen-mat}.
The $OD_{\pm}5$-brane and the $OD3$-brane only can satisfy
eq.\eq{odgen-mat} and eq.\eq{odgen+mat}, respectively.
As was discussed in eq.\eq{dynod-}, the $OD_{\pm}5$-brane cannot
preserve the supersymmetry $q_+^-$ in this case. On the other
hand, since the term, $\partial_\sigma X^{\rd}$, for the $OD3$-brane
is absent in eq.\eq{dynq-sigma}, the $OD3$-brane does not belong to the
present consideration but does to the previous case 3.3.2.
Thus any dynamical supersymmetry of
$OD$-branes is not preserved under the general background.

\section{Supersymmetric D-branes with $F_{IJ}$}

In this section we will study supersymmetric boundary conditions
to preserve dynamical supersymmetries after turning on the flux
$F_{IJ}$ \ct{matt}. As we mentioned, the gluing matrix in this case is
given
by eq.\eq{magOmega}. We can also derive the
conservation law for the dynamical supersymmetry in eq.\eq{dyn-susy}
with the gluing matrix $\O$ in eq.\eq{magOmega}
\be \la{con-law-mag}
\frac{\pa q^-_\tau}{\pa \tau} + \frac{\pa q^-_\sigma}
{\pa \sigma} = 0.
\ee
The current $q^-_\sigma$ at the boundary is reduced to
\bea \la{ds-magcurrent}
q^-_\sigma|_{\pa \S} &=& \sqrt{\frac{2}{p^+}}
\Bigl[ \pa_\t X^{r^\prime} \gamma^{r^\prime} S^1 -
\half \Bigl(\pa_\s X^{r} - \Bigl(\frac{1-N}{1+N} \Bigr)^{rs} \pa_\t
X^{s} \Bigr)(\delta^{rt} + N^{rt}) \gamma^t S^1 \xx
&& + \frac{\mu}{2} X^r \gamma^s (N^{rs} + \delta^{rs} \Gamma_B) \Omega \Pi S^1
- \frac{\mu}{2} X^{r^\prime} \gamma^{r^\prime} (1 - \Gamma_B) \Omega \Pi S^1
\Bigr],
\eea
where we defined
\bea \la{def-triality}
&& \O \ga^r \O^T = - N^{rs} \ga^s,\\
\la{def-magGamma}
&& \G_B = \Pi \O^T \Pi \O^T.
\eea
In eq.\eq{ds-magcurrent}, we already used the fermionic boundary
condition \eq{bc-fermion} and the relation \eq{comm-gaom}.
Note that $N^{rs} = \delta^{rs}$ when $F_{IJ}=0$ and then we recover
eq.\eq{ds-current}.

Here we have taken a different recipe from Mattik's \ct{matt}.
Indeed we only assumed the fermionic boundary
condition \eq{bc-fermion}, whose form is independent of the detail of
backgrounds, to get the result \eq{ds-magcurrent}. The relations
\eq{comm-gaom} and \eq{def-triality} are the direct consequences
(Baker-Campbell-Hausdorff formula) of the prescribed form of
the gluing matrix $\O$ in eq.\eq{magOmega}.
We will now find most general boundary conditions which give rise to the
vanishing current at the boundary, i.e., $q^-_\sigma|_{\pa \S}=0$.

It is convenient to divide the Neumann directions into two groups:
$r=(a,i)$ where $a,b,c$ denote the directions without flux and $i,j,k$
denote those with flux. We also introduce a sign flip operation
$ \pi: X^{1,2,3,4} \mapsto  - X^{1,2,3,4}$.
In this notation, eq.\eq{def-triality} can be solved as follows:
\bea \la{triality-flux}
&& \exp^{\frac{1}{4} \Theta_{jk} \ga^{jk}} \ga^i \exp^{-\frac{1}{4}
  \Theta_{jk} \ga^{jk}} = - N^{ij} \ga^j, \\
\la{triality-0}
&& N^{ab} = \delta^{ab}.
\eea
Also the matrix $\G_B$ can be rewritten as follows
\be
\G_B = \widetilde{\G} \exp^{-\frac{1}{4} \pi(\Theta_{jk}) \ga^{jk}} \exp^{-\frac{1}{4}
\Theta_{jk} \ga^{jk}},
\ee
where $\widetilde{\G}= \Pi \widetilde{\O} \Pi \widetilde{\O}$.

Using the above results, eq.\eq{ds-magcurrent} can be separably written into two parts:
\bea \la{magcurrent2}
q^-_\sigma|_{\pa \S} &=& \sqrt{\frac{2}{p^+}}
\Bigl[ \Bigl(\pa_\t X^{r^\prime} \gamma^{r^\prime}
- \frac{\mu}{2} X^{r^\prime} \gamma^{r^\prime} (1 - \Gamma_B) \Omega
\Pi - \pa_\s X^{a} \gamma^a + \frac{\mu}{2} X^a \gamma^a (1 + \Gamma_B)
\Omega \Pi \Bigl) S^1 \xx
&& - \half \Bigl(\pa_\s X^{i} + {\cal F}^{ij} \pa_\t X^{j} \Bigr)(\delta^{ik}
+ N^{ik}) \gamma^k S^1 + \frac{\mu}{2} X^i \gamma^j (N^{ij} + \delta^{ij}
\Gamma_B) \Omega \Pi S^1 \Bigr],
\eea
where we defined
\be \la{def-F}
 {\cal F}^{ij} = - \Bigl(\frac{1-N}{1+N} \Bigr)^{ij}.
\ee
Looking into the terms in eq.\eq{magcurrent2}, we see that the most
pertinacious term is the last one, which is related to the Neumann coordinates
and cannot be cancelled with other terms due to its peculiar form.
So we have to demand \ct{matt} that
\be \la{crucialmatrix-flux}
N^{ij}  \gamma^j + \gamma^i \Gamma_B = 0.
\ee
The above equation can be satisfied if and only if
\bea \la{gamma=1}
&& \widetilde{\G}= \Pi \widetilde{\O} \Pi \widetilde{\O} = 1, \\
\la{gamma-flux}
&&  \ga^i \exp^{-\frac{1}{4} \pi(\Theta_{jk}) \ga^{jk}} = \exp^{\frac{1}{4}
\Theta_{jk} \ga^{jk}} \ga^i, \;\;\; \forall i.
\eea
The condition \eq{gamma-flux} is equivalent to
\be \la{flip-flux}
\pi(\Theta_{jk}) =  \Theta_{jk} \;\; \mbox{and} \;\;
\mbox{rank}(\Theta_{jk}) = 2.
\ee

We see that the coordinates $X^i$ in the limit $\Theta_{jk} = 0$ where $N^{ij}
= - \delta^{ij}$ satisfy the usual Dirichlet boundary condition $\pa_\t X^{i}
= 0$ while in the limit $\Theta_{jk} = \pi$ where $N^{ij}
= \delta^{ij}$ they satisfy the usual Neumann boundary condition $\pa_\s X^{i}
= 0$. The conditions \eq{gamma=1} and \eq{flip-flux} thus say that we have to start from
a $D_+$-brane when $\Theta_{jk} = 0$ and the magnetic flux should be extended along
only two directions in $X^{1,2,3,4}$ or $X^{5,6,7,8}$ to have a
D-brane to preserve the dynamical supersymmetry. So the D-brane
with magnetic flux is continuously interpolating from a
$D_+$-brane to a  $D_-$-brane with, in general, different amount
of supersymmetries at the endpoints. We will see that the
dynamical supersymmetry can be preserved by the same amount as
$D_+$-branes
only if the condition \eq{crucialmatrix-flux} is satisfied.
So the maximally supersymmetric cases are
$\widetilde{\O} = 1, \, \Pi, \, \ga \Pi$ which correspond to $(+,-,0,0), \,
(+,-,4,0),\, (+,-,0,4)$ branes when  $\Theta_{jk} = 0$. These are exactly the
cases found by Mattik \ct{matt}.

Under the condition \eq{crucialmatrix-flux}, the current in
eq.\eq{magcurrent2} is reduced to
\bea \la{magcurrent}
q^-_\sigma|_{\pa \S} &=& \sqrt{\frac{2}{p^+}}
\Bigl[ \pa_\t X^{r^\prime} \gamma^{r^\prime} S^1 -
\pa_\s X^{a} \gamma^a S^1 + \mu X^a \cos\frac{\Theta_{jk}}{2} \gamma^a \widetilde{\Omega} \Pi
S^1 \xx
&& - \mu X^{r^\prime} \sin{\frac{\Theta_{jk}}{2}} \gamma^{r^\prime} \ga^{jk}
\widetilde{\Omega} \Pi S^1
- \half \Bigl(\pa_\s X^{i} + {\cal F}^{ij} \pa_\t X^{j} \Bigr)(\delta^{ik}
+ N^{ik}) \gamma^k S^1 \Bigr].
\eea
Now it is easy to find bosonic boundary conditions to preserve the dynamical
supersymmetry. First of all, we have the following boundary conditions
\bea \la{dbc-flux}
&& \pa_\t X^{r^\prime} = 0 =  x_0^{r^\prime},\\
\la{nbc-flux}
&& \pa_\s X^{i} + {\cal F}^{ij} \pa_\t X^{j} = 0.
\eea
For $\widetilde{\O} = 1$, $X^a = 0$ by definition, so that the dynamical
supersymmetry is maximally preserved. For $\widetilde{\O} = \Pi$ and
$\ga \Pi$, $\widetilde{\Omega} \Pi S^1 = S^1$ so that the supersymmetry
is maximal if
\be \la{4004}
\pa_\s X^{a} - \mu \cos(\half \Theta_{jk}) X^a  = 0.
\ee
This is the same kind of the boundary condition
for the $(+,-,4,0), \, (+,-,0,4)$ branes.

For the other branes, the dynamical supersymmetry can also be preserved by
considering the spinor satisfying $\widetilde{\Omega} \Pi S^1 = \pm S^1$ at
the boundary, but this time only 4 dynamical supersymmetries are preserved
as was shown in \ct{jhep} since the projection operator
$\half(1 \pm \widetilde{\Omega} \Pi)$ is now nontrivial.
This case also requires the modified Neumann boundary condition
like eq.\eq{mbc-d+} with the replacement $\mu \rightarrow \mu \cos(\half
\Theta_{jk})$. Note that the matrix $\widetilde{\Omega}$ corresponding
to the $(+,-,1,1)$ and $(+,-,2,2)$ branes only can satisfy the condition
\eq{crucialmatrix-flux} since the $(+,-,3,3)$ and $(+,-,4,4)$ branes
cannot have additional Neumann directions satisfying eq.\eq{flip-flux}.
Note that the brane position can be arbitrary when $\Theta_{jk} = 0$.

\section{Supersymmetric D-branes in general background}

Now we will relax the condition \eq{dbc-flux}. First note that, as
shown in the previous section, the projected spinors defined by
\be \la{pro-spinor}
S^A_{\pm} \equiv \half(1 \pm \widetilde{\Omega} \Pi) S^A
\ee
can only preserve the dynamical supersymmetry in the magnetic flux
background. So we have to consider the spinor $S^1_{\pm}$
satisfying
\be \la{mag-flux}
\gamma^{a r^\prime jk} S^1_\pm = \pm S^1_\pm
\ee
or
\be \la{mag-boost}
\gamma^{r^\prime s^\prime jk} S^1_\pm = \pm S^1_\pm,
\ee
where the $\pm$ sign in the right-hand side is just an eigenvalue
of the matrix $\gamma^{a r^\prime jk}$ or $\gamma^{r^\prime s^\prime
jk}$ (independent of that in eq.\eq{pro-spinor})
and we will not concern the sign.

For the case \eq{mag-flux}, we have to further introduce a
constant electric flux generated by the linear gauge field $A_+ =
- F_{+a} X^a$ in addition to the quadratic piece eq.\eq{mbc-d+}, where
$F_{+a} = \mu x_0^{r^\prime} \sin{\frac{\Theta_{jk}}{2}}$. In this
case the Neumann boundary condition is given by
\be \la{el-mag}
\pa_\s X^{a} - \mu \cos \frac{\Theta_{jk}}{2} X^a
+ \mu x_0^{r^\prime} \sin{\frac{\Theta_{jk}}{2}} = 0.
\ee
The D-brane with $\widetilde{\O} = 1$ does not belong to the above case
since $X^{a} = 0$ by definition. However, we can introduce two
independent fluxes for the D-branes with $\widetilde{\O} = \Pi$
and $\ga \Pi$ - at least 2 dynamical supersymmetries.
For example, $\gamma^{a r^\prime jk} = \ga^{1756}$ or $\ga^{2856}$
for $\O = \Pi \exp^{\frac{1}{2} \Theta_{56} \ga^{56}}$.
The other cases allow only one independent projection, so that
they also preserve 2 dynamical supersymmetries.
For example, $\gamma^{a r^\prime jk} = \ga^{1237}$ or $\ga^{1248}
= - \ga \widetilde{\O} \Pi \ga^{1237}$
for $\O = \ga^{3456} \exp^{\frac{1}{2} \Theta_{12} \ga^{12}}$.

For the case \eq{mag-boost}, on the other hand, we need the
modified Dirichlet boundary condition
\be \la{boost-mag}
\pa_\t X^{\pr} - \mu v^{\ps} \sin \frac{\Theta_{jk}}{2} = 0,
\ee
where $v^{\ps} \equiv X^{\ps}|_{\pa \S}$ for some $\ps \in D$.
This means that the D-brane is constantly boosted along the
$\pr$-direction. When $\Theta_{jk} = 0$, it satisfies the usual
Dirichlet boundary condition, consistent with 3.2.
The D-brane with $\widetilde{\O} = 1$ can have two
independent boosts (or angular momenta) while the D-branes with $\widetilde{\O} = \Pi$
and $\ga \Pi$ allow only one angular momentum, so that
they preserve at least 4 dynamical supersymmetries. The other
branes can have only one independent boost, so at least 2
dynamical supersymmetries are preserved.

In order to consider eq.\eq{mag-flux} and eq.\eq{mag-boost}
simultaneously, we need at least three Dirichlet directions and
$X^a \neq 0$. This is satisfied only by the brane, for example,
with $\O = \ga^{35} \exp^{\frac{1}{2} \Theta_{12} \ga^{12}}$.
This brane can preserve 2 dynamical supersymmetries since
$\gamma^{a r^\prime jk} = \ga^{1236}$ and $\gamma^{\ps t^\prime jk}
= \ga^{1278} = - \ga \widetilde{\O} \Pi \ga^{1236}$.

\section{Discussion}

We studied D-branes in the type IIB plane wave background together
turning on additional backgrounds - electric as well as magnetic
fluxes and an angular momentum. We found a much richer spectrum of
supersymmetric D-branes than in the flat spacetime. Let us briefly
summarize the results obtained in this paper.

It turned out that the $D_-$-branes and the $D_+$-branes behave
very differently when an electric flux and an angular momentum are
turned on. The $D_-$-branes can be placed away from the origin by
introducing a constant electric flux. So this process in general
breaks a global world-volume symmetry except some special case.
For a $(+,-,m,n)$ brane, the D-brane worldvolume theory has the
global symmetry $SO(m) \times SO(n) \times SO(4-m) \times SO(4-n)$.
The breaking of the symmetry $SO(m)$ or $SO(n)$ by the electric
flux is necessarily correlated with that of $SO(4-m)$ or $SO(4-n)$
due to the shift of transverse position. One exception is a
$(+,-,3,1)$ or $(+,-,3,1)$ brane which preserves the maximal
supersymmetry as discussed in 3.1.1. In this case the electric
flux and the transverse shift do not touch the global symmetry
$SO(3) \times SO(3)$. $D_+$-branes, however, do not break any
global symmetry by the electric flux. But, in this case, the
electric flux is not constant but linearly proportional to Neumann
coordinates. Note that the $D_+$-branes can take arbitrary
transverse position without breaking supersymmetry \ct{jhep}. We also
observed that the $D_-$-branes can also move with constant
velocity preserving some amount of supersymmetries. However the
$D_+$-branes cannot move while preserving the supersymmetry.

Since the oblique D-branes contain both $D_-$-like and $D_+$-like
supercharges, a similar feature also appears as the $D_-$-branes.
But, only if the electric flux is turned on to preserve the
$D_+$-like supercharge, some OD-branes can then be shifted away
from the origin after further introducing a constant electric flux
or move with constant velocity, while preserving some
supersymmetries.

We also considered the magnetic flux background. As observed by
Mattik \ct{matt}, we showed that the D-brane with magnetic flux is continuously
interpolating from a $D_+$-brane to a  $D_-$-brane, in general,
with different amount of supersymmetries at the endpoints. Our
analysis shows that the OD-branes cannot preserve any dynamical
supersymmetry in the magnetic flux background. We observed that
there exist
supersymmetric moving D-branes satisfying eq.\eq{boost-mag}. Note that
these D-branes already carry the electric flux $F_{+a}$ as well as
the magnetic flux $F_{jk}$. These D-branes thus correspond to
giant gravitons rotating in the $X^{\pr}-X^{\ps}$ plane with the nontrivial
worldvolume gauge field. For example, the D-brane with
$\O = \ga^{35} \exp^{\frac{1}{2} \Theta_{12} \ga^{12}}$
can preserve 2 dynamical supersymmetries with the nontrivial
$F_{+3,5},\; F_{12}$ and $L_{78}$ background. Recently this kind of
giant graviton was found \ct{seok} in the $AdS_5 \times S^5$
geometry. It could be interesting to see whether the giant
graviton in \ct{seok} after the Penrose limit can be reduced to that in the
plane wave geometry.

In this paper we did not consider intersecting D-branes
\ct{bdl,ohta}. It should be straightforward to extend the analysis
in this paper to the intersecting D-branes following the scheme in
\ct{cha}.

\section*{Acknowledgments}

HSY thanks Tako Mattik for a helpful correspondence.
This work was supported by the Korean Research Foundation Grant
KRF 2003-015-C00111 (BHL), KRF 2003-070-C00011 (CP and JwL),
the Basic Research Program of the Korea Science and Engineering
Foundation Grant No. R01-2004-000-10526-0 (BHL) and
the Research Institute for Basic Science of Sogang University (CP).
The work of HSY is supported by the Alexander von Humboldt
Foundation and he also thanks the staff of the Institut f\"ur Physik,
Humboldt Universit\"at zu Berlin for the cordial hospitality.

\appendix

\section{Matrix relations}

Here we list useful matrix relations for the
$OD$-branes with $\G = \ga^{1256}$ in table \ref{tableone}
which were used to find supersymmetric backgrounds in the
subsection 3.3.

$\bullet \; OD3$-brane with $\O = \frac{1}{2}(\gamma^{1}-\gamma^{6})
(\gamma^{2} + \gamma^{5})$:
\bea \la{0d3+}
&& \O \Pi P_+ = \ga^{2345} P_+ \; \mbox{or} \; -\ga^{1346} P_+, \\
\la{0d3-}
&& \O \Pi P_- = -\ga^{34} P_- \; \mbox{or} \; - \ga \ga^{78} P_-.
\eea

$\bullet \; OD5$-brane with $\O = \frac{1}{2}(\gamma^{1}-\gamma^{6})
(\gamma^{2} - \gamma^{5}) \ga^{34}$:
\bea \la{0d5+34}
&& \O \Pi P_+ = P_+, \\
\la{0d5-34}
&& \O \Pi P_- = \ga^{25} P_- \; \mbox{or} \; \ga^{16} P_-.
\eea

$\bullet \; OD5$-brane with $\O = \frac{1}{2}(\gamma^{1}-\gamma^{6})
(\gamma^{2} - \gamma^{5}) \ga^{78}$:
\bea \la{0d5+78}
&& \O \Pi P_+ = - \ga P_+, \\
\la{0d5-78}
&& \O \Pi P_- =  \ga \ga^{25} P_- \; \mbox{or} \; \ga \ga^{16} P_-.
\eea

$\bullet \; OD5$-brane with $\O = \frac{1}{2}(\gamma^{1}-\gamma^{6})
(\gamma^{2} + \gamma^{5}) \ga^{37}$:
\bea \la{0d5+37}
&& \O \Pi P_+ = - \ga^{2457} P_+ \; \mbox{or} \; \ga^{1467} P_+ \\
\la{0d5-37}
&& \O \Pi P_- =  - \ga^{47} P_- \; \mbox{or} \; -\ga \ga^{38} P_-.
\eea

$\bullet \; OD7$-brane with $\O = \frac{1}{2}(\gamma^{1}-\gamma^{6})
(\gamma^{2} + \gamma^{5}) \ga^{3478}$:
\bea \la{0d7+}
&& \O \Pi P_+ = - \ga^{2578} P_+ \; \mbox{or} \; -\ga^{1678} P_+ \\
\la{0d7-}
&& \O \Pi P_- = \ga^{78} P_- \; \mbox{or} \; \ga \ga^{34} P_-.
\eea

\newpage

\bibliographystyle{JHEP-like}

\end{document}